\documentclass{svproc}
\usepackage{url}

\usepackage{graphicx}
\usepackage{multicol}
\usepackage{footmisc}
\usepackage{amsmath,amssymb,amsfonts}
\usepackage{xcolor}%
\usepackage{lineno}

\begin{document}
\mainmatter              
\title{Establishing Earth's Matter Effect in Atmospheric Neutrino Oscillations at IceCube DeepCore}
\titlerunning{Establishing Earth's Matter Effect}  
%
\author{Anuj Kumar Upadhyay\footnote[2]{also at Institute of Physics, Sachivalaya Marg, Sainik School Post, Bhubaneswar 751005, India and Department of Physics, Aligarh Muslim University, Aligarh 202002, India.}  \\
(For the IceCube Collaboration\footnote[1]{\url{http://icecube.wisc.edu}})}
\authorrunning{Anuj Kumar Upadhyay (For the IceCube Collaboration)} 
%
\tocauthor{For the IceCube Collaboration}
\institute{Dept. of Physics and Wisconsin IceCube Particle Astrophysics Center, University of Wisconsin-Madison, Madison, WI 53706, USA \\
\email{aupadhyay@icecube.wisc.edu}}

\maketitle              

\begin{abstract}
The discovery of the non-zero value of $\theta_{13}$ has opened an exciting opportunity to probe the Earth's matter effects in three-flavor oscillations of atmospheric neutrinos. These matter effects depend on both neutrino energy and the electron density distributions encountered during their propagation through Earth. In this contribution, we present preliminary sensitivities from the DeepCore detector, a densely instrumented sub-array of the IceCube neutrino observatory at the South Pole, demonstrating its ability to observe these matter effects in atmospheric neutrino oscillations. Using simulated data equivalent to 9.3 years of observations at IceCube DeepCore, we show the sensitivity of the DeepCore to reject the vacuum oscillation hypothesis and align with the Preliminary Reference Earth Model. Additionally, we present the expected improvement in sensitivity for rejecting the vacuum oscillations using the upcoming IceCube Upgrade, a low-energy extension of the IceCube detector.

\end{abstract}
\section{Introduction}\label{sec:intro}
Neutrino oscillations can be explained by expressing the neutrino flavor eigenstates ($\nu_e$, $\nu_\mu$, $\nu_\tau$) as a superposition of their mass eigenstates ($\nu_1$, $\nu_2$, $\nu_3$), governed by three mixing angles ($\theta_{12}$, $\theta_{13}$, $\theta_{23}$) and a CP-violating Dirac phase, $\delta_\text{CP}$. The neutrino oscillations also depend on the mass-squared differences, $\Delta m^2_{21}$ and $\Delta m^2_{31}$. In the past twenty years, most of the neutrino oscillation parameters have been measured with a precision of a few percentages, except for the three unknowns: (i) the value of $\delta_\text{CP}$, (ii) the octant of $\theta_{23}$, and (iii) the neutrino mass ordering. The precisely measured oscillation parameters, especially the discovery of a non-zero value for $\theta_{13}$ by the Daya Bay experiment~\cite{DayaBay:2012fng}, have opened up the possibility of exploring the matter effects experienced by the upward-going atmospheric neutrinos as they pass through Earth.

Atmospheric neutrinos are produced when cosmic rays interact with nuclei in Earth's atmosphere, resulting in muon-flavor and electron-flavor neutrinos, as well as their antiparticles. They traverse a wide range of baselines, from 20 km to approximately 13000 km, and cover energy ranges from a few MeV to hundreds of TeV. At energies below about 15 GeV, both flavors of neutrinos exhibit distortions in their oscillation probabilities as they transverse through Earth due to coherent forward scattering with ambient electrons. The resonant enhancement in oscillation probability is known as the Mikheyev-Smirnov-Wolfenstein (MSW) resonance. Additionally, neutrinos passing through Earth's core experience a sharp density jump at the core-mantle boundary, which leads to significant changes in their oscillation probabilities. This effect is known as the parametric resonance or neutrino oscillation length resonance. These resonances depend on both the neutrino energy and the electron density distribution encountered as they travel through Earth.

The IceCube Neutrino Observatory~\cite{IceCube:2016zyt} is a cubic kilometer ice-Cherenkov detector, located at the geographic South Pole. It consists of 5160 Digital Optical Modules (DOMs), deployed on 86 vertical strings, with a photomultiplier tube as the main component to detect photons. These DOMs capture Cherenkov light emitted by secondary charged particles produced in neutrino interactions with ice, as well as from atmospheric muons. The bottom central region of the IceCube detector is known as the DeepCore sub-array which comprises 8 densely packed strings. The energy threshold for the densely instrumented DeepCore volume is a few GeV, while for IceCube it is about 100 GeV.

In this study, we analyze simulated low-energy atmospheric neutrino data in the energy range of approximately 5 GeV to 100 GeV for the IceCube DeepCore detector equivalent to 9.3 years of observations. We aim to estimate the sensitivity of the IceCube DeepCore detector to observe Earth's matter effects in atmospheric neutrino oscillations by rejecting the vacuum oscillation hypothesis with respect to that with matter oscillations based on the Preliminary Reference Earth Model (PREM)~\cite{Dziewonski:1981xy}. At present, we use simulated data to demonstrate the robustness of the analysis. Later, we will perform this analysis with real experimental data.

\section{Event sample}
\label{sec:reconstruction}

In this analysis, we use a Monte Carlo (MC) simulated event sample corresponding to 9.3 years of observations at the IceCube DeepCore detector. This MC sample has a significantly improved signal-to-background ratio, with the background less than 1\% of the sample. The primary background in DeepCore consists mainly of atmospheric muons and random detector noise. In this sample, properties of interacting neutrinos, such as energy, arrival direction, and particle identification (PID), are reconstructed using a Convolutional Neural Networks (CNNs) based machine learning algorithm. A PID classifier is employed to distinguish between track-like topologies, primarily produced by $\nu_\mu$ charged-current (CC) interactions, and cascade-like topologies, which result from $\nu_e$ CC and all neutral-current interactions. Additionally, DeepCore is also sensitive to $\nu_\tau$ CC interactions, which contribute to both track-like and cascade-like topologies. Details about the background filtering process and the reconstruction of neutrino observables can be found in Refs.~\cite{IceCubeCollaboration:2023wtb,IceCube:2024xjj}.

We binned the final level simulated MC events in terms of reconstructed energy, reconstructed $\cos\theta_\text{zenith}$, and PID. The MC events are re-weighted based on Earth's density models (PREM or vacuum profile) and the systematic uncertainty parameters considered in the analysis. These systematic uncertainties are related to the atmospheric neutrino flux, cross sections, detector response, and neutrino oscillation parameters, and are treated as nuisance parameters in the analysis. The details of these systematic uncertainty parameters can be found in Refs.~\cite{IceCubeCollaboration:2023wtb,IceCube:2024xjj}.

\section{Sensitivity results}
\label{sec:results}

This section presents the Asimov sensitivity for establishing the Earth's matter effects in atmospheric neutrino oscillations by rejecting the vacuum oscillations with respect to the matter oscillations. Rejecting the vacuum oscillation hypothesis with respect to the PREM profile is a binary hypothesis test, where the Asimov sensitivity is defined according to Ref.~\cite{Ciuffoli:2013rza}. The Asimov sensitivity is shown as a function of the true value of $\sin^2\theta_{23}$, given that $\theta_{23}$ remains most uncertain parameter apart from $\delta_\text{CP}$.
\begin{figure}[h!]
	\centering
	\includegraphics[width=0.495\linewidth]{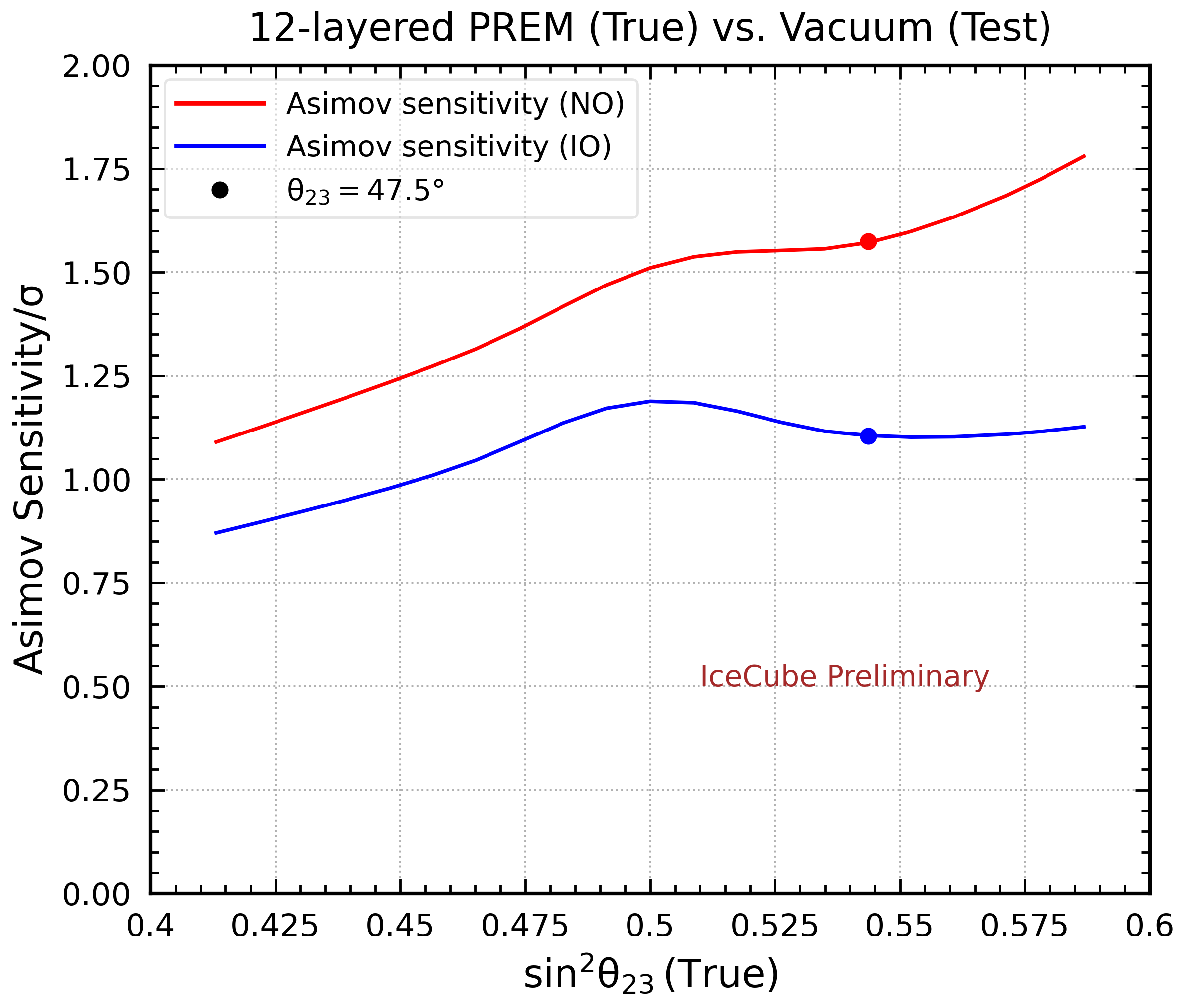}
	\includegraphics[width=0.495\linewidth]{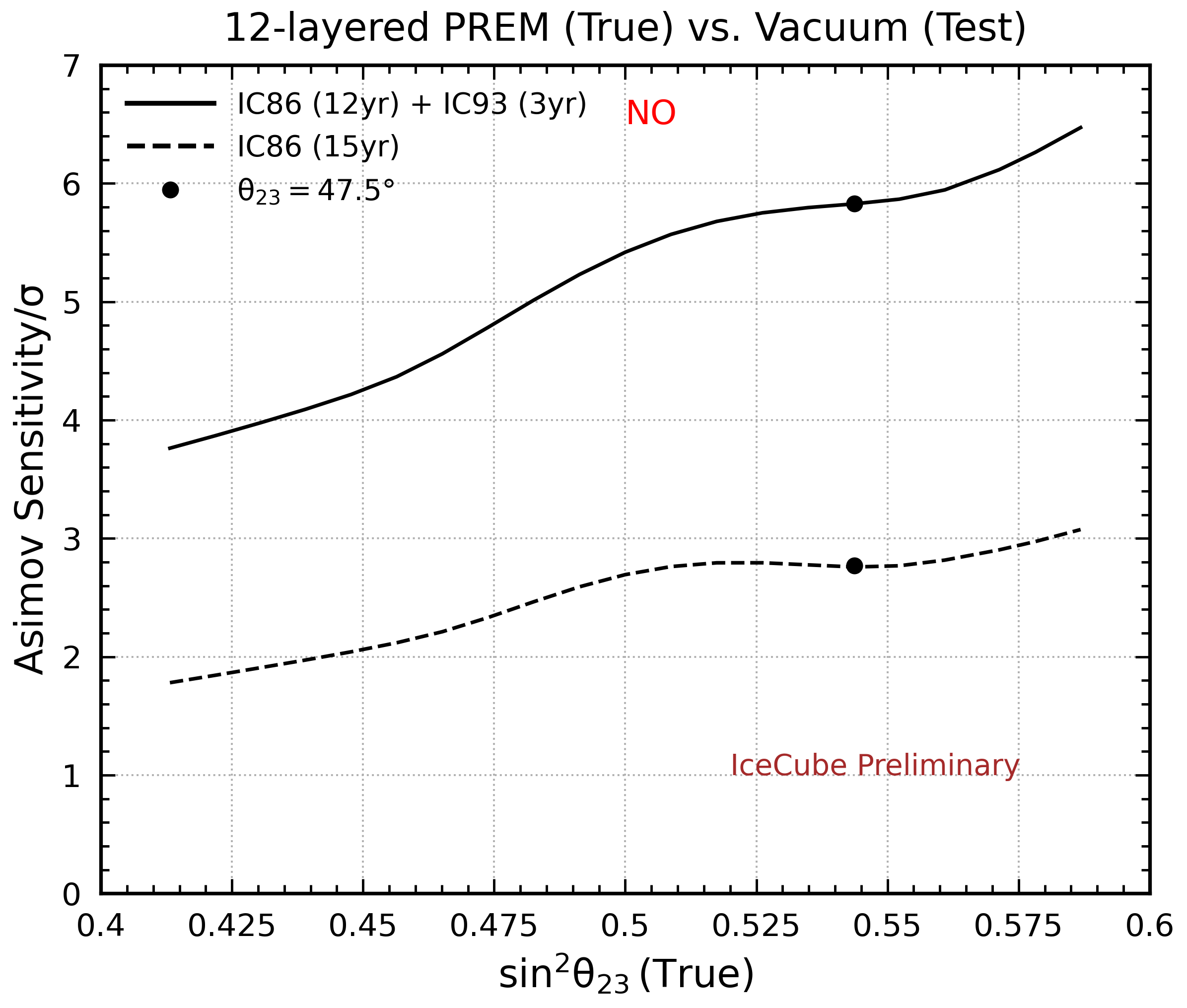}
	\caption{Asimov sensitivity as a function of the true value of $\sin^2\theta_{23}$ to establish the Earth's matter effects in atmospheric neutrino oscillations. The left panel shows the expected sensitivity based on IceCube DeepCore simulated data, equivalent to 9.3 years of observations, while the right panel presents the expected sensitivity using the current DeepCore setup only (dashed line) and with the Upgrade included along with DeepCore (solid line).}	
	\label{fig:results}
\end{figure} 
Since $\delta_\text{CP}$ has minimal impact on this analysis, we kept it fixed at the true value of $\delta_\text{CP}=0$. The other oscillation parameters, $\theta_{13}$, $\theta_{12}$, and $\Delta m^2_{21}$ are also fixed, as they have been precisely measured. During the fit, $\Delta m^2_{31}$ and $\theta_{23}$ are kept free along with other nuisance parameters. 

In the left panel of Fig.~\ref{fig:results}, the red and blue curves correspond to the Asimov sensitivity assuming normal and inverted neutrino mass ordering (NO and IO), respectively, in both simulation (true) and theory (test). These curves are estimated using IceCube DeepCore simulated data equivalent to 9.3 years of observations. The red and blue dots mark the Asimov sensitivity for a representative true value of $\theta_{23} = 47.50^\circ$, yielding sensitivities of $1.57 \sigma$ for NO and $1.10 \sigma$ for IO, respectively. 

The right panel of Fig.~\ref{fig:results} demonstrates the expected sensitivity for two scenarios, both assuming NO. The first scenario, represented by the dotted curve and labeled ``IC86 (15yr)'', considers the current setup of the IceCube DeepCore with continuous data-taking. The second scenario, shown by the solid curve and labeled ``IC86 (12yr) + IC93 (3yr)'', combines simulated data equivalent to 12 years of IC86 observations with an additional 3 years of simulated data from the IceCube Upgrade, which includes seven extra strings starting in 2026. The IceCube Upgrade setup is expected to provide improved sensitivity.

\section{Conclusion}
\label{sec:conclusion}

We presented the preliminary Asimov sensitivity for observing the Earth's matter effects in atmospheric neutrino oscillations using the IceCube DeepCore simulated data equivalent to 9.3 years of observations. Additionally, we showcased the potential of the upcoming IceCube Upgrade~\cite{IceCube:2023ins} detector, which is expected to improve sensitivity to matter effects by approximately three times due to its lower energy threshold and improved systematic uncertainties.

\subsubsection{\ackname} We acknowledge the financial support from the Department of Atomic Energy (DAE), Govt. of India and the INSPIRE Ph.D. fellowship provided by the Department of Science and Technology (DST), Govt. of India.
	

\bibliographystyle{spphys}
\bibliography{References}

\begin{thebibliography}{1}
\providecommand{\url}[1]{{#1}}
\providecommand{\urlprefix}{URL }
\expandafter\ifx\csname urlstyle\endcsname\relax
  \providecommand{\doi}[1]{DOI \discretionary{}{}{}#1}\else
  \providecommand{\doi}{DOI \discretionary{}{}{}\begingroup
  \urlstyle{rm}\Url}\fi

\bibitem{DayaBay:2012fng}
F.P. An, et~al., Phys. Rev. Lett. \textbf{108}, 171803 (2012).
\newblock \doi{10.1103/PhysRevLett.108.171803}

\bibitem{IceCube:2016zyt}
M.G. Aartsen, et~al., JINST \textbf{12}(03), P03012 (2017).
\newblock \doi{10.1088/1748-0221/12/03/P03012}.
\newblock [Erratum: JINST 19, E05001 (2024)]

\bibitem{Dziewonski:1981xy}
A.M. Dziewonski, D.L. Anderson, Phys. Earth Planet. Interiors \textbf{25}, 297
  (1981).
\newblock \doi{10.1016/0031-9201(81)90046-7}

\bibitem{IceCubeCollaboration:2023wtb}
R.~Abbasi, et~al., Phys. Rev. D \textbf{108}(1), 012014 (2023).
\newblock \doi{10.1103/PhysRevD.108.012014}

\bibitem{IceCube:2024xjj}
R.~Abbasi, et~al., arXiv e-prints  (2024).
\newblock \doi{10.48550/arXiv.2405.02163}

\bibitem{Ciuffoli:2013rza}
E.~Ciuffoli, J.~Evslin, X.~Zhang, JHEP \textbf{01}, 095 (2014).
\newblock \doi{10.1007/JHEP01(2014)095}

\bibitem{IceCube:2023ins}
P.~Eller, et~al., PoS \textbf{ICRC2023}, 1036 (2023).
\newblock \doi{10.22323/1.444.1036}

\end{thebibliography}

\end{document}